\documentclass[twoside]{article}[12pt, a4paper]

\usepackage{graphicx}
\usepackage[a4paper,top=2.5cm,bottom=2.5cm,left=3cm,right=3cm,marginparwidth=1.75cm]{geometry}
\usepackage{authblk}
\usepackage{graphicx} 
\usepackage[dvipsnames]{xcolor}
\usepackage{array}
\usepackage{tabularx}
\usepackage{booktabs}
\usepackage{float}
\usepackage{subcaption}
\usepackage{parskip}
\usepackage{setspace}

\usepackage{fancyhdr}
\title{LLM-based event abstraction and integration for IoT-sourced logs}
\author[1,2]{Mohsen Shirali}
\author[3]{Mohammadreza Fani Sani}
\author[1]{Zahra Ahmadi}
\author[1]{Estefanía Serral}

\affil[1]{Research Centre for Information Systems Engineering (LIRIS), KU Leuven, Brussels 1000, Belgium (\texttt{zahra.ahmadi,estefania.serralasensio\}@kuleuven.be})}
\affil[2]{ICTeam, UCLouvain, Louvain-la-Neuve, Belgium 
\texttt{mohsen.shirali@uclouvain.be}}
\affil[3]{Microsoft Development Center Copenhagen, Copenhagen, Denmark (\textit{mfanisani@microsoft.com}}

\date{}

\begin{document}

\maketitle
{\small{Corresponding author: Mohsen Shirali (\texttt{mohsen.shirali@uclouvain.be})}}

\begin{abstract}

The continuous flow of data collected by Internet of Things (IoT) devices, has revolutionised our ability to understand and interact with the world across various applications. However, this data must be prepared and transformed into event data before analysis can begin. In this paper, we shed light on the potential of leveraging Large Language Models (LLMs) in event abstraction and integration. Our approach aims to create event records from raw sensor readings and merge the logs from multiple IoT sources into a single event log suitable for further Process Mining applications. We demonstrate the capabilities of LLMs in event abstraction considering a case study for IoT application in elderly care and longitudinal health monitoring. The results, showing on average an accuracy of 90\% in detecting high-level activities
. These results highlight LLMs' promising potential in addressing event abstraction and integration challenges, effectively bridging the existing gap.

\vspace{24pt}
\textbf{Keywords. Internet of Things \and Large Language Models  \and Event Abstraction \and Log Integration \and Multi-modality.}
\end{abstract}

\section{Introduction}
\label{sec:intro}

The IoT technology has the potential to create a new "cyber-physical" world, where "things" can directly operate, act, and influence the physical world ~\cite{karkouch2016data}. Smart devices have seamlessly integrated into everyday life, and low-cost sensors are embedded in various applications, from smart homes to smart cities and industrial IoT in smart factories.
To extract meaningful insights from raw IoT data, advanced data mining techniques are required. However, raw sensor-level data is unstructured and non-informative, making it unsuitable for data mining. Raw data must be transformed into event logs to make the information discoverable and comprehensible for analysis~\cite{senderovich2016road}. Without proper pre-processing, the analysis foundation is compromised, creating an abstraction gap~\cite{sani2020preprocessing,diba2020extraction}.

Looking in particular into Process Mining (PM), as one of the data analysis disciplines, it requires event logs, where each event represents a single activity at a particular time~\cite{jessen2023chit}. However, in complex settings, systems often lack process-centric data, rendering the data unsuitable for immediate analysis. Likewise, in smart homes, events are often logged at the sensor trigger level, which is too granular for meaningful pattern mining, complicating the application of PM techniques~\cite{diba2020extraction}.
Additionally, the required data often resides in various databases and/or generated by multiple information systems or sources. Data from multiple sources must be integrated to provide a comprehensive view, facilitating the understanding of system behaviour and enhancing analysis capabilities from various angles.

Preparing event logs is typically manual, requiring domain knowledge to group correlated tasks into meaningful activities. This manual effort can be inconsistent and error-prone, especially for complex processes involving multiple data sources~\cite{FaniSani2023LLM}. Moreover, integrating heterogeneous logs introduces challenges like format inconsistencies, timestamp misalignment, and data quality issues~\cite{jia2017big}. Besides, event data may have mixed granularity and contextual information, complicating the application of data analysis techniques~\cite{tax2018mining}.

In this way, pre-processing techniques are vital to build meaningful event logs before any actual analysis, like PM, can begin~\cite{diba2020extraction} and resources have to be allocated for this purpose.
A study in~\cite{Stein2024Loss} revealed that the laborious effort for data preparation or the necessity for proper expertise can terminate PM projects and cause them to be called off. If event logs are not generated properly, the potential benefits of using recorded logs for data analysis are significantly limited.

Log abstraction and integration are two essential data preparation or pre-processing tasks often used in data-driven analytics, to enhance result interpretability~\cite{van2021event}.
Furthermore, log abstraction involves summarising or grouping low-level data elements into higher-level event-based representations, aligning with the executed activities in a process.
This simplifies the analysis and assigns meaning to sets of raw data records, allowing them to be interpreted as a single event~\cite{rebmann2022multi,diba2020extraction}.
Log integration also merges combining data from multiple sources/systems to create a unified view for analysis~\cite{Liu2024Turning}.

Existing event abstraction techniques are either unsupervised or supervised~\cite{de2020event,van2021event}. Unsupervised methods rely on control-flow similarities between low-level event types without requiring input on targeted high-level activities~\cite{rebmann2022multi}. Conversely, supervised methods require extensive input on high-level activities, demanding significant manual effort and domain knowledge~\cite{rebmann2022multi}.
These two extremes necessitate a balance to support users in abstraction, addressing the challenges of manual effort, domain knowledge, and consistency. 

Motivated by these challenges, we propose using Large Language Models (LLMs) to automate raw data pre-processing and event log generation, minimising the need for user background knowledge and effort. LLMs, trained on extensive data, are ideal for this task, capable of processing textual information and performing natural language understanding and generation tasks~\cite{Berti2024Abstraction}. Our approach aims to utilise LLMs to analyse IoT sensor logs, generate meaningful event records, and merge logs from multiple sources, facilitating Process Mining applications.

We explore how an LLM, given a prompt with event label samples and minimal user input, can abstract sensor readings into event logs. In the prompt, we outline the task, clarify the role of the LLM, describe the inputs it will process, and specify the expected outputs. Hence, the contributions of our work are as follows; first, We use LLMs to automate the detection and labelling of low-level sensor data to create abstracted events. Second, we develop a method for abstracting and generating events based on data streams, making the proposed approach suitable for online applications. Third, information from multiple sources will be merged into a unified event log to enhance analytics by providing a comprehensive view of the processes.
Our work aims to streamline the process of event log generation, making it more efficient and accessible, and ultimately improving the utility of IoT-sourced sensor data in data-driven analytical applications.

The rest of this article is organised as follows:
Section~\ref{sec:background} reviews the existing event abstraction and integration methods and describes the multi-source IoT dataset which is used for evaluation. Section~\ref{sec:llm} describes the proposed LLM-based event log abstraction and integration approach.
Then, in Section~\ref{sec:results_andEvaluation}, the results and performance of the proposed approach are evaluated and a discussion on the usefulness of utilising LLMs besides IoT systems is provided in Section~\ref{sec:discussion}. 
Finally, Section~\ref{sec:conclusion} concludes the article's findings.
\vspace{-0.2cm}
\section{Background knowledge}
\label{sec:background}
\vspace{-0.2cm}

\subsection{Existing works for event abstraction and integration}
\label{sec:RW}
The journey from raw data to event logs suitable for PM has been extensively studied in works like~\cite{diba2020extraction} and~\cite{van2021event}.
For example, van Zelst \textit{et al.}~\cite{van2021event} and Fahland \textit{et al.}~\cite{Fahland2022Extracting} have used techniques like log abstraction to group events into higher-level activities for simplified visualisation and analysis
Additionally, a hierarchical framework for event abstraction based on notion of activity instances was proposed in~\cite{li2021activity}. Similarly, Senderovich \textit{et al.}~\cite{senderovich2016road} suggested a knowledge-driven approach to transform raw sensor data into standardised event logs.
The integration of Complex Event Processing (CEP) and BPM, as highlighted by Soffer \textit{et al.} has gained attention for creating more efficient systems, especially for IoT applications.
Mangler~\textit{et al.}~\cite{mangler2024internet} also address the challenges of connecting process data with IoT data and propose a semi-automatic framework for transforming low-level IoT sensor data into higher-level process events.
Furthermore, Di Federico and Burattin~\cite{Federico2023CvAMoS} introduce CvAMoS, a method to identify recurring sequences of activities and consider their context of execution for event abstraction.
Additionally, \cite{Seiger2023Interactive} introduces the concept of activity signatures for deriving knowledge about activity executions and training supervised learning models to detect higher-level process activity executions for larger datasets.

Moreover, it is particularly useful (or even crucial in many use cases) to build a ground truth based on domain knowledge and user expertise in the early stages of IoT data analysis, which can later facilitate automated activity detection~\cite{Seiger2023Interactive}. For instance, in some IoT logs, human expertise can help in interpreting sensor-level data and mapping them to activity-level events using samples, without predefined descriptions.This approach, used in activity recognition, maps sensor readings such as opening the door of a refrigerator to activities like "preparing meal"~\cite{tax2018mining}. 
In addition, system designers can translate sensor readings into events using predefined rules. For example, \cite{Shirali2020Mobility} and \cite{Lull2021behaviour} used senor locations to create event logs for discovering behaviour models and mobility patterns by PM techniques. Also, studies like~\cite{falah2022probabilistic,Cook2013CASAS} consider sensor types and attached objects to create event logs for performed activities of daily living (ADLs).

To overcome the required domain knowledge,~\cite{rebmann2022multi} applies a method to learn event representations and automatically suggest related event groups for abstraction. This approach allows users to select meaningful event groups without initial input. Additionally, an unsupervised method in~\cite{rebmann2023unsupervised} continuously transforms user interactions into event streams for downstream analysis.
Another unsupervised learning technique is also proposed in~\cite{Janssen2021} to discretize sensor data into identifiable activities, and refine their sequences by deducing sub-models to identify concurrency and interleaving within the data in smart home scenario.
Furthermore, recent studies have explored using LLMs for event abstraction like \cite{FaniSani2023LLM} that grouped tasks into high-level activities and assigned labels to them based on their similarity.

In terms of log integration methods to address data heterogeneity, in~\cite{VanEck2015PM}, logs from different organizational departments were combined for end-to-end process analysis. Also, in~\cite{sarno2017hierarchy}, a time-based heuristics miner is used to discover high-level and low-level process models in parallel from multi-source logs, integrating them with PetriNet refinement.

\vspace{-0.3cm}
\subsection{Multi-source IoT Dataset}
\label{sec:case_study}
\vspace{-0.2cm}
Our study aimed to assess the effectiveness and accuracy of using LLMs on IoT-sourced datasets to create event logs suitable for Process Mining.
We applied our proposed LLM to sensor logs from an IoT dataset collected in an Ambient Assisted Living scenario.
This dataset contains 146 days of data, capturing the daily activities of a 60-year-old woman living independently in an apartment~\cite{falah2022probabilistic}. The data was collected using ambient sensors installed in the house, a wristband, and a smartphone.

\begin{figure}[b]
    \vspace{-0.3cm}
    \centering
    \includegraphics[width=0.80\textwidth, height = 5 cm]{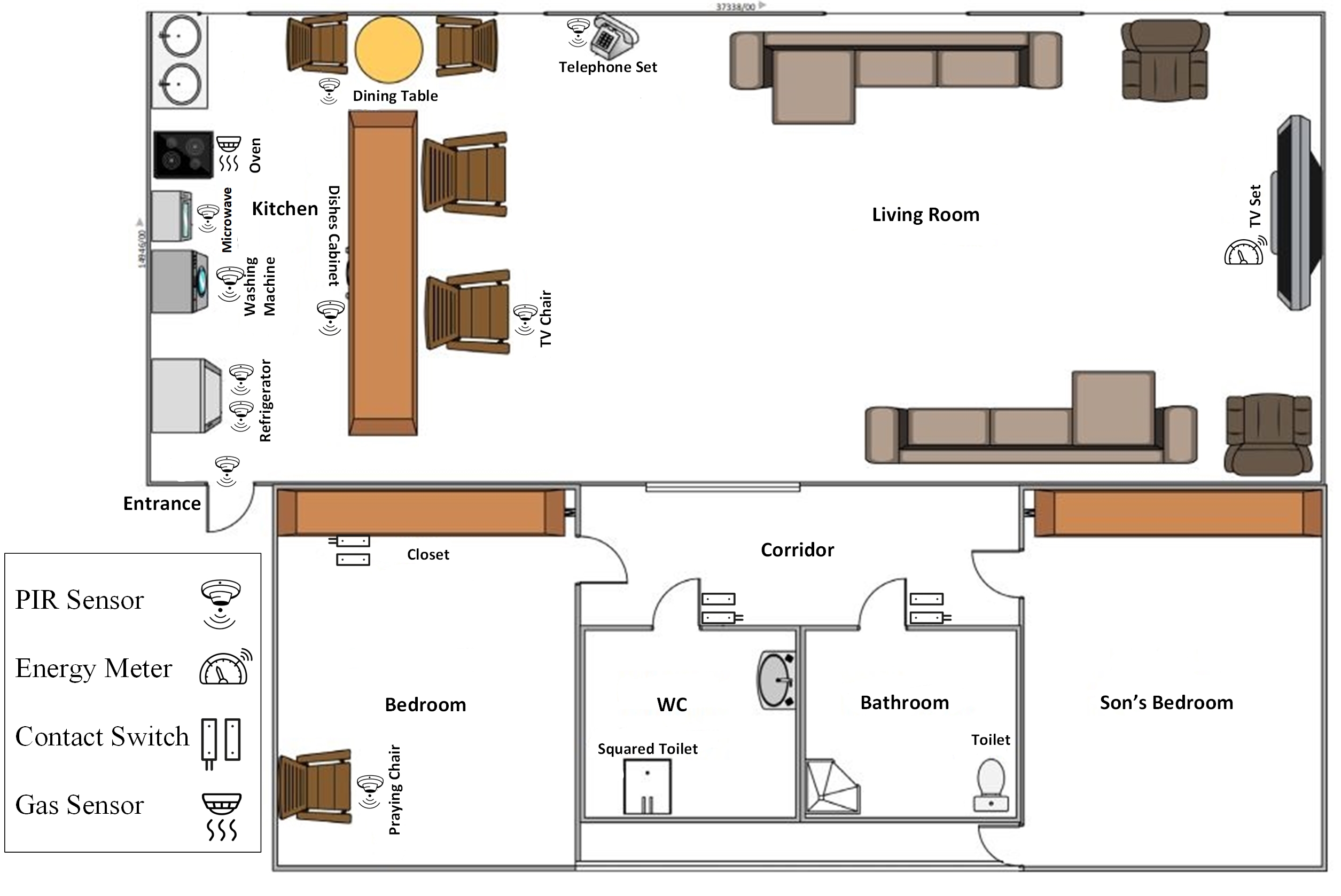}
    \vspace{-0.2cm}
    \caption{The house floor plan, location and types of ambient sensors}
    \label{fig1-homeplan}
    \vspace{-0.4cm}
\end{figure}

The ambient sensors included PIR sensors placed on furniture and appliances, a power usage sensor indicating TV usage, contact sensors for detecting door openings and closings, and a gas detection sensor for identifying cooking activity.
These sensors were positioned in six different areas of the house, as shown in Figure~\ref{fig1-homeplan} and recorded event data with timestamps, sensor names, and sensor values (e.g., on/off states).
The participant also wore a Xiaomi Mi Band-3 wristband to collect sleep-related information such as sleep time, duration, and quality.
Additionally, smartphone usage data, including timestamps and the names of used applications, were collected using an application.

The ambient sensors, such as the TV sensor and kitchen appliance sensors, are triggered when the participant performs specific activities.
As a result, their measurements are at a low level of abstraction representing the presence of the subject near the appliances, the open or closed state of doors, and the use of the stove.
By knowing the timestamp of their triggered states, we can discover the corresponding activities to express the exact start and end times and the duration of the conducted activity.
These activities are mainly known as instrumented Activity Daily Livings (iADL), the activities performed using specific instruments and are investigated in several behaviour modelling and analysis studies and projects, like~\cite{Cook2013CASAS,shirali2024synergy}.


With the entire sequence of captured records for a specific duration, we can infer events related to the person’s presence in different areas of the house, the performed activities, and the sequence of their movements.
For this purpose, an event abstraction step is needed to elevate the abstraction level of sensor raw readings into recognisable activity events in an event log.
The events related to smartphone usage should also be converted and inserted into the log, while the names of applications can be replaced with their type or with a general label such as `Using Smartphone' to preserve the participants' privacy. 
Furthermore, the sleep-related information collected by the wristband will complete the preferred event log. 

\begin{figure}[bt]
     \centering
     \begin{subfigure}[h]{\textwidth}
         \centering
         \includegraphics[width=\textwidth]{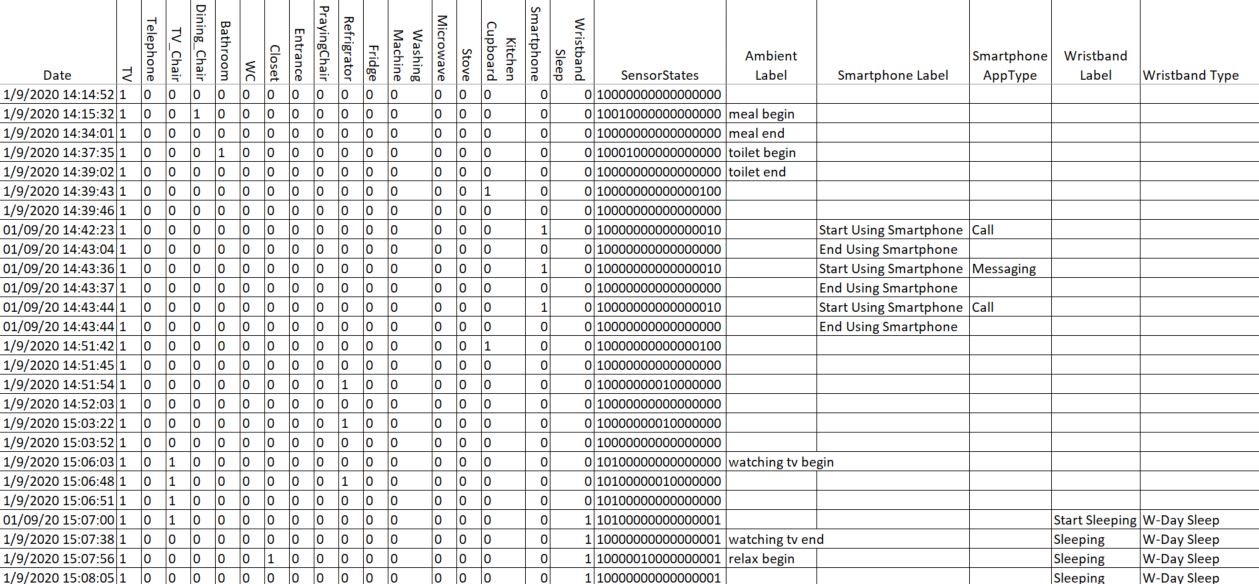}
         \caption{We have 17 sensors and three categories of labels.}
         \label{Fig:SensorLog}
     \end{subfigure}
     \begin{subfigure}[h]{0.75\textwidth}
         \centering
         \includegraphics[width=\textwidth]{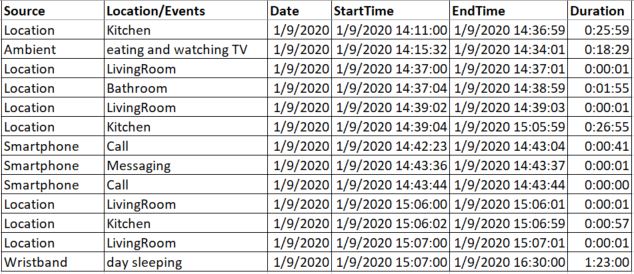}
         \caption{Activities and presence in locations can be concurrent and be happened on different days.}
         \label{Fig:EventLog}
     \end{subfigure}
     \vspace{-0.1cm}
     \caption{A view of (a) sensor logs and (b) resulted event log (the ground truth version) for the experimented IoT dataset.}
     \label{Fig:sensor_event_log}
     \vspace{-0.4cm}
\end{figure}

In the current case study, the event log includes location events, such as  \textit{Bedroom}, \textit{Bathroom}, \textit{Kitchen},
and the activity events like \textit{sleeping}, \textit{meal preparation}, \textit{praying}, {watching tv}.
A snapshot of sensor logs and the required event log, including the information of all modalities (the ground truth version), is depicted in Figure~\ref{Fig:sensor_event_log}.
We fed these sensor logs obtained from devices' reports in our multi-modal IoT dataset into the proposed LLM model, asking it to create a PM-friendly event log.
The resulting event log was then compared with a ground truth event log that was generated through a pre-processing step by applying multiple event detection rules determined by domain experts and then inspecting manually to correct any falsely detected events, as described and used in~\cite{shirali2024synergy}.

\vspace{-0.3cm}

\section{Event log abstraction and integration by LLMs} 
\label{sec:llm}
\vspace{-0.2cm}
As previously stated, PM techniques rely on having an event log as an input, and without it, gaining insights is not possible.
While many scientific works assume the availability of the event log, the procedure of pre-processing raw data, detecting events, abstracting them into activities, and generating an event log can be time-consuming in real scenarios. Abstracting events from sensor data, particularly when dealing with streams of input data and potential concurrent events and activities recorded by a combination of sensors, can be quite challenging~\cite{mangler2024internet}.
In this section, we explain how we can use Generative AI and LLMs to (1) abstract the events from sensor data to activities, and (2) generate an event log from different resources. 

\subsection{Abstraction of events into activities}
We propose to use LLMs to detect activities from sensor data. In this regard, we give some basic explanations about the sensors, possible activity labels, and a few examples to LLM to provide us with a label for the detected change in the status of sensors. 
In other words, we call an LLM whenever we have a change in sensor values.
The changes in the binary values of the sensors have the highest significance regarding process-level events, as it is stated in the~\cite{mangler2024internet}.
Thus, each sensor log entry should denote a change in the status of sensors.
Therefore, we can consider this task as a classification task in that LLM will be the classifier and activity labels are the classes.
The number of classes is the number of possible desired activities plus one as we may encounter sensor changes that do not correspond to any specific activity. However, there are instances where sensor changes occur without any corresponding activities due to noises, irrelevant sensors, etc. In such cases, a blank class is defined to address these situations.

For this task, we utilised GPT-4 as the model and applied few-shot learning~\cite{yong2023prompt} and chain of thought~\cite{wei2022chain} techniques.
In the prompt template, we incorporated placeholders for the label of activities, the sensor explanation, and the output format\footnote{The sources and examples for our implemented LLM with more detailed information are publicly available at https://github.com/mfanisani/LLM4IoT}. 



In general, the prompt can be modified if we have more information based on the rules that we have in business.
For instance, here the output of all sensors in our dataset are binary values (i.e., 0 and 1), and it is important to note that we take this advantage to save tokens and reduce the cost by just passing the aggregated sensor information in a binary format to LLM (the number of consumed tokens in a generated a prompt is one of the limitations in using LLMs). Hence, the LLM will need to recognise which value corresponds to which sensor based on the order of binary digits in the sensor state. Some examples of SensorStates are given in Figure~\ref{Fig:SensorLog}.
We provide the SensorStates for the current and previous state in our prompt to simplify the task for the LLM. The LLM will first identify what has changed in these two values and then provide the activity name.


\subsection{Integration and event log generation}

We can have several information sources for one event log. 
Note that for each activity we have two events; the first one corresponds to the start of the activity and the second one indicates that the activity has ended.
In real world, we may have more events for each activity considering the complete life cycle of activities~\cite{bernardi2016activity}. 
We propose to use LLM to gather information from different resources and combine them into one event log.
However, we have some limitations: 
\begin{itemize}
    \item We cannot give the whole information to LLM as we have a limitation on the number of tokens,
    \item We do not have a caseID\footnote{caseID is one of the main columns in any event log that represents the process instance.} as all the information related in our dataset is related to one process instance.
\end{itemize}

To overcome the above limitations, we considered date as caseID. This will allow us to analyse the process of activities that have been done each day.
Therefore, our data will include the date as the case ID, the activity, start time, and end time.
However, it is possible that an activity start on one day and end on the next day, e.g. sleeping from 2022-11-23 22:00 to 2022-11-24 07:00.
In this case, we will provide LLM with the information for activities that span two days and ask it to trim the event at midnight, and adjust the event log for each day accordingly.
To handle this, we introduce another column with binary values indicating whether the activity is completed on the same date or not. If the activity crosses over to the next day, we will not include the end time for that next day.
While we assumed activities span at most two consecutive days to mitigate the constraint on the number of tokens per LLM call, it's important to recognise that real-world scenarios may not adhere to this assumption and may have longer execution times.


Our prompt template includes possible activities, two-shot examples, the information for the date of interest and the following day, and the format of the desired output. 
In this way, we can increase the performance of the event log generation task, as we will call LLM once per day. 
Note that the output does not contain the header of event log columns.
We use a basic script to collect the output information generated by LLM for different dates and compile it into a single event log.  

The proposed framework for event log abstraction and integration can be used on different sensors or data logs with different prompts and on various LLMs. The samples for the appropriate labels are given to the model and it returns the expected output for the stream of data inputs immediately.
Therefore, the framework is also able to deal with streams of data sources and the output of this layer is sufficient for the PM tools without any further involvement of the users and interventions.

\vspace{-0.2cm}
\section{Evaluation}
\label{sec:results_andEvaluation}
It is important to assess the effectiveness of LLMs in event abstraction by analysing the results and evaluating the model's accuracy in generating event records and assigning the appropriate labels. This section presents the results of proposed LLM-based event abstraction on the real-world IoT-sourced dataset for our case study.
We have conducted the evaluation in two steps. First, the abstracted events for each data source are examined and compared with the intended labels to measure the accuracy of the LLM model in identifying and abstracting the raw sensor records for each data source. This comparison will highlight the LLM model's potential in dealing with different IoT sources.
In the next step, the outcome of merging three logs from ambient, smartphone and wristband into the final integrated PM-friendly event log is cross-checked with the ground truth event log.  

\vspace{-0.2cm}
\subsection{Event abstraction and integration results}
\label{sec:results}
The accuracy results of the LLM in the abstraction of events foreach of the three modalities in our dataset are provided in Table~\ref{tab:results}.
For ambient sensors, we just consider activities with an occurrence of at least 50 times. The 21 labels in this source correspond to 10 activities with a start and end plus a label that corresponds to no activity.  
For smartphones and wristbands, we have respectively 13 and 3 labels indicating the type of used application or day/night sleep.
For each activity, we have start, continue and end labels. Similar to ambient sensors, we have a label for no activity. 

The accuracy for ambient sensor data is much lower than the other datasets as the number of related sensors and the number of labels are higher. 
For all three datasets, the lowest accuracy is for \textit{None} label that corresponds to no activity. In several cases, we predicted an activity (e.g., \textit{Watching TV} start/end) instead of \textit{None}.  
We randomly checked some of the wrongly assigned labels and in almost all of them, the reasoning part produced by LLM was wrong.

\begin{table}[b]
    \vspace{-0.5cm}
    \caption{The accuracy of our activity detection method for different sensors} \label{tab:results}
    \centering
    \begin{tabular}{p{0.3\textwidth}p{0.25\textwidth}p{0.18\textwidth}p{0.18\textwidth}}
    \toprule
    & \textbf{Ambient sensors} & \textbf{Smartphone} & \textbf{wristband} \\
    \cmidrule{2-4}
    \textbf{Number of labels} & \multicolumn{1}{c}{21} & \multicolumn{1}{c}{13} & \multicolumn{1}{c}{3} \\
    \textbf{Accuracy} & \multicolumn{1}{c}{0.81} & \multicolumn{1}{c}{0.97} & \multicolumn{1}{c}{0.92} \\
    \bottomrule
    \end{tabular}
    \vspace{-0.2cm}
\end{table}
\vspace{-0.1cm}


\vspace{-0.2cm}
\subsection{Evaluation of LLM-generated event log}
\label{sec:evaluation}
The evaluation of the LLM-based event log in terms of the accuracy of the recognised events is crucial to assess the success of LLMs in the abstraction and integration task.
This evaluation involves comparing two event logs that contain various events with their respective start and end times.
A snapshot of the event log generated using LLM is presented in Table~\ref{tab:snapshot}.
Therefore, the matching of events labels, events orders, and the duration and time precedence of events are the important factors to be considered for evaluation.
We have utilised the number of correctly generated events, the ratio of perfectly aligned date and Edit Distance Alignment (EDA)~\cite{fani2020conformance} metrics to measure the success rate of our proposed LLM-based approach.





\begin{table}[tb]
\centering
\vspace{-0.2cm}
\caption{The snapshot of the LLM-generated event log from three sources (the values of Date column are caseIDs).}
\label{tab:snapshot}
\scriptsize
\begin{tabular}{|l|l|l|l|l|l|}
\hline
Date     & Type       & Activity  & Start Time & End Time & Next date \\ \hline
1/8/2020 & Ambient    & sleeping  & 23:04:33   & 08:28:59 & True      \\ \hline
1/8/2020 & Ambient    & toilet    & 23:05:39   & 23:07:21 & False     \\ \hline
1/8/2020 & Smartphone & messaging & 23:17:28   & 23:17:49 & False     \\ \hline
1/8/2020 & Smartphone & messaging & 23:17:52   & 23:18:31 & False     \\ \hline
1/8/2020 & Smartphone & messaging & 23:18:32   & 23:20:28 & False     \\ \hline
1/8/2020 & Smartphone & call      & 23:20:32   & 23:20:45 & False     \\ \hline
$\cdot\cdot\cdot$ & $\cdot\cdot\cdot$ & $\cdot\cdot\cdot$      & $\cdot\cdot\cdot$   & $\cdot\cdot\cdot$ & $\cdot\cdot\cdot$     \\ \hline
\end{tabular}
\vspace{-0.2cm}
\end{table}

In our experiment, we provided the GPT4-0613 model with sensor states and their original labels, tasking it to generate an event log. Notably, we used the ground truth labels, not those predicted by LLM. The ground truth dataset contains 17,165 events, while the LLM-generated event log has 15,420 events. Among these generated events, 12,741 were correctly detected, with 11,365 having matching start and end activities.
In addition, LLM is able to generate all the unique activities that are defined for it.  

In our analysis, we observed that self-loops—consecutive events with the same label—play a significant role. After removing these self-loops, we found that the generated event log contains 11,248 events, while the ground truth dataset has 11,431 events. Interestingly, LLM sometimes aggregates self-loops, particularly when the user interacts with her smartphone, where the interactivity times between self-loops are negligible.

To show how the generated event log is aligned with the ground truth event log, we applied EDA between similar dates in both event logs using the technique proposed in~\cite{fani2020conformance}.
This metric measures how similar the sequences of events are, where a value closer to \textit{one} indicates higher similarity.  
The results are presented in Table~\ref{tab:alignment} and indicate that the event log generated by LLM captures similar behaviour compared to the ground truth event log but it aggregated the self-loops.

\begin{table}[tb]
    \vspace{-0.2cm}
    \caption{Behavioral comparison of event logs with/without considering self loops.}
    \label{tab:alignment}
    \centering
    \begin{tabular}{p{0.15\textwidth}p{0.26\textwidth}p{0.15\textwidth}p{0.26\textwidth}}
    \toprule
    \multicolumn{2}{c}{\textbf{Edit Distance Alignment}} & \multicolumn{2}{c}{\textbf{Perfectly Aligned Dates(\%)}} \\
    \midrule
    \textbf{All events} & \textbf{Without Self-loops} & \textbf{All events} & \textbf{Without Self-loops} \\
    \multicolumn{1}{c}{0.83} & \multicolumn{1}{c}{0.94} & \multicolumn{1}{c}{0} & \multicolumn{1}{c}{33}\\
    \bottomrule
    \end{tabular}
    \vspace{-0.2cm}
\end{table}


\vspace{-0.4cm}
\section{Discussion}
\label{sec:discussion}
The results of this case study indicate that utilising LLMs can significantly alleviate the challenges associated with event log abstraction and integration for IoT-sourced logs. LLMs can bridge the gap between data collection and the demand for event logs at appropriate abstraction levels for data analysis techniques, including Process Mining. Although, traditional methods for event abstraction and integration are often manual, error-prone, and time-consuming, but, LLMs can automate and speed up these tasks while reducing human errors. They can swiftly pre-process large volumes of raw sensor data, identify and abstract events with high accuracy, and uniformly label them, thus freeing up human resources for more strategic tasks.
However, it’s important to acknowledge that the output of our proposed solution may not always be highly reliable. Therefore, we recommend incorporating human oversight in applications that require a high level of reliability.

Key advantages of using LLMs in event log generation and integration for IoT systems include:

\begin{itemize}
    \item \textbf{Automation and efficiency.} LLMs can automate the pre-processing of raw data and make the process more efficient by reducing manual effort and errors. 
    
    \item \textbf{Reduced need for domain knowledge.} Expertise is crucial in supervising the abstraction process and interpreting IoT logs requires substantial domain knowledge to map low-level sensor readings to high-level activities. LLMs, trained on diverse datasets, can understand and abstract sensor data into meaningful events without deep domain-specific knowledge. This allows more users with limited background knowledge to generate valuable insights and make IoT data analytics more feasible.

    \item \textbf{Handling multi-modality.} IoT systems often collect data from multiple sources, resulting in mixed granularity and inconsistencies in data formats and characteristics. LLM-based models can understand and integrate heterogeneous data inputs, by handling this complexity. These models can align data from various sensors, ensuring a cohesive and consistent event log, particularly for scenarios involving time misalignment and varied data structures.

    \item \textbf{Real-time processing.} LLMs are capable of handling streams of data inputs, making them suitable for online applications. Initially, in our study, we triggered LLM calls whenever there was a change in sensor values. However, this approach can be computationally expensive and slow in certain applications. To address this limitation, we propose using batching—calling the LLM fewer times by grouping input data— similar to our approach for event log generation. This means that LLMs can process incoming data in real-time, handling sensor records one by one without relying on maintaining extensive change histories, and continuously updating the event logs with new information. This ensures that the event logs are always current, facilitating timely analysis and decision-making.

    \item \textbf{Performance optimisation with prompt engineering and user feedback.} Designing prompts that clearly describe the task, input data, and expected output, can significantly enhance the quality of the LLM's output.
    Moreover, users' and domain experts involvement in the process remains crucial for fine-tuning the LLM's outputs and improving accuracy over time. A continuous feedback loop between LLMs and users helps in refining the model, adapting to evolving data patterns, and ensuring that the system meets the specific needs of the application. This collaboration between humans and intelligent systems ensures that the benefits of LLMs are fully realised while maintaining high standards of accuracy and reliability.
\end{itemize}

In the end, LLMs offer a powerful solution to the challenges of event log abstraction and integration in IoT systems. By automating the manual, repetitive, and error-prone data preparation tasks, LLMs streamline the process, reduce the need for extensive domain knowledge, and handle the complexities of multi-modality and mixed granularity in data. Additionally, with proper prompt engineering and user feedback mechanisms, LLMs can provide accurate and up-to-date event logs, enhancing the overall efficiency and effectiveness of IoT data analytics. Hence, integrating LLMs with IoT systems can revolutionise the way we process and analyse sensor data, unlocking new potential for real-time insights and decision-making.
\vspace{-0.4cm}

\section{Conclusion and future works}
\label{sec:conclusion}
The rapid advancement of IoT technologies promises valuable data-driven insights and enhanced operational efficiency. However, a gap exists between the raw IoT data and the processed data needed for analysis, hindering the full potential of IoT technologies. In this study, LLMs are utilised to aid users in pre-processing raw data, converting it into event logs containing all the essential fields for event data analytics techniques, like Process Mining. The results underscore LLMs' potential in addressing event log abstraction and integration for IoT-generated logs. Despite these promising outcomes, the use of LLMs in this field is in its early stages, with room for improvement. Future research could focus on applying LLMs to text-based raw data, improving prompt engineering techniques, and fine-tuning LLMs to further enhance their performance and usefulness in data preparation.


        




%
%

\subsubsection*{Acknowledgement.} Mohsen Shirali, is now affiliated at UCLouvain in Belgium under grant Win4Collective number 2310088. The work of Zahra Ahmadi and Estefanía Serral was supported by the Flemish Fund for Scientiﬁc Research (FWO) with grant number G0B6922N.

%
%
%
\bibliographystyle{splncs04}
\bibliography{ref}

\end{document}